\documentclass[12pt]{article}

\usepackage{amsmath}
\usepackage{amssymb}
\usepackage{amsfonts}
\usepackage{latexsym}
\usepackage{color}

\catcode `\@=11 \@addtoreset{equation}{section}

\catcode `\@=12

\newcommand{\be}{\begin{equation}}
\newcommand{\en}{\end{equation}}
\newcommand{\bea}{\begin{eqnarray}}
\newcommand{\ena}{\end{eqnarray}}
\newcommand{\beano}{\begin{eqnarray*}}
\newcommand{\enano}{\end{eqnarray*}}
\newcommand{\bee}{\begin{enumerate}}
\newcommand{\ene}{\end{enumerate}}

\newcommand{\mc}{\mathcal}

\newcommand{\F}{{\cal F}}

\newcommand{\Lc}{{\cal L}}
\newcommand{\C}{{\cal C}}
\newcommand{\1}{1 \!\! 1}

\newcommand{\Hil}{\mc H}

\catcode `\@=11 \@addtoreset{equation}{section}
\catcode `\@=12

\begin{document}

\thispagestyle{empty}

\vspace*{2cm}

\begin{center}
{\Large \bf  Extended pseudo-fermions from non commutative bosons}\\[10mm]

{\large S. Twareque Ali} \footnote[1]{Department of Mathematics and Statistics, Concordia University,
Montr\'eal, Qu\'ebec, CANADA H3G 1M8\\
e-mail: twareque.ali@concordia.ca}
\vspace{3mm}\\

{\large F. Bagarello} \footnote[2]{ Dipartimento di Energia, ingegneria dell'Informazione e modelli Matematici,
Facolt\`a di Ingegneria, Universit\`a di Palermo, I-90128  Palermo, ITALY\\
e-mail: fabio.bagarello@unipa.it\,\,\,\, Home page: www.unipa.it/fabio.bagarello}
\vspace{3mm}\\

{\large Jean Pierre Gazeau}\footnote[3]{Laboratoire APC, Universit\'e Paris  Diderot, Sorbonne Paris-Cit\'e, 10, rue A. Domon et L. Duquet,
75205 Paris Cedex 13, France, and Centro Brasileiro de Pesquisas F\'isicas,
 Rio de Janeiro, 22290-180 Rio de Janeiro, Brasil\\e-mail: gazeau@apc.univ-paris7.fr}
\end{center}

\vspace*{2cm}

\begin{abstract}
\noindent We consider some modifications of the two dimensional canonical commutation relations, leading to {\em non commutative bosons} and we
show how biorthogonal bases of the Hilbert space of the system can be obtained out of them. Our construction extends those recently introduced by one of us (FB), modifying the
canonical anticommutation relations. We also briefly discuss how bicoherent states, producing a resolution of the identity, can be defined.

\end{abstract}


\vfill


\newpage

\section{Introduction}

In a recent series of papers the possibility of modifying the canonical (anti)commutation relations in order to get biorthogonal (Riesz) bases has been
considered in some detail, see \cite{bagpb5}, \cite{bagpf} and references therein. The articles \cite{bagpb7}-\cite{abg} are more concerned with physical applications. The
functional structure arising from these modifications turns out to be rather rich and, moreover, it appears to be closely related to the
so-called non hermitian quantum mechanics (or some variations of it). This particular aspect is discussed, for instance, in \cite{bagpb9,bagpb10}.

\vspace{2mm}

On the different side, a rather fashionable topic in the recent literature is the so called {\em noncommutative quantum mechanics} which, in the form that is of relevance to us, is
essentially a two dimensional version of the position and momentum operators, $\hat x_j$ and $\hat p_j$, which satisfy, other than $[\hat x_j,\hat
p_k]=i\,\delta_{j,k}$, $j=1,2$, also the commutation rule $[x_1,x_2]=i\theta$, for some real parameter $\theta$ (see \cite{ncqm1,ncqm2}). These were originally adopted to describe a possible quantized space at very small length scales.  {For recent coherent state and group theoretical approaches to non commutative quantum mechanics see, for example, \cite{ncqmjpg09,ncqmjpg11,chowali}. {In \cite{chowali} it has been shown that the commutation relations of non commutative quantum mechanics are essentially those satisfied by the generators of the Galilei group in (2+1) space-time dimensions with two central extensions, a fact which had earlier been also noted in \cite{horduvsticht}. Additionally, using the coherent states introduced in  \cite{chowali}, these same commutation relations have been derived by the method of coherent state quantization.}

In this paper we will show how these two topics could be indeed related, and how a non commutative version of the annihilation and creation
operators acting on a Hilbert space $\Hil$ naturally gives rise to two biorthonormal bases of $\Hil$. More in details, after a first no-go
result, showing that, not surprisingly, not all the non commutative {\em extensions} of two-dimensional quantum mechanics gives rise to these
kind of bases, we discuss a second possibility for which several interesting facts can be established. In particular, two biorthogonal sets are
explicitly found, together with several extended versions of non self-adjoint number-like operators of the same kind as those introduced in
\cite{bagpf} {(In a related paper \cite{balshali} such biorthogonal bases have been realized as biorthogonal sets of polynomials in a complex
variable)}. These non self-adjoint operators are related to their adjoints by certain intertwining operators, which can be easily identified in
our setting. This is the content of Section II, while in Section III we give an explicit realization of the construction. In Section IV we
return to the general construction for the infinite-dimensional Hilbert space, and in Section V we construct related coherent states. Section
VI contains our conclusions.

\section{Generalized pseudo-fermions}

We will discuss in this section how non commutative quantum mechanics can be used to {\em generate}  biorthogonal bases of an infinite dimensional Hilbert space which arise from a numerable union of other, finite-dimensional, pairs of biorthogonal bases, each pair in some finite dimensional Hilbert space. We will also describe in detail the intertwining operators relating, see below, the number-like operators which naturally arise in our framework. Moreover, since these latter are not self-adjoint, we will briefly discuss the natural connections between our results and non hermitian quantum mechanics.

We begin our analysis with a simple but useful exercise, showing that not all the non commutative deformations of ordinary two-dimensional quantum mechanics give rise to an interesting structure, when analyzed from the point of view of \cite{bagpb5} and \cite{bagpf}.

\subsection{A no-go result}

Consider the following transformation $(x,y,p_x,p_y)\rightarrow (X,Y,P_X,P_Y)$, where \be \left\{
    \begin{array}{ll}
        X=x-\frac{\theta}{2}\,p_y,\quad P_X=p_x  \\
        Y=y+\frac{\theta}{2}\,p_x,\quad P_Y=p_y,\\
       \end{array}
        \right.
\label{21}\en which maps the canonical operators $(x,y,p_x,p_y)$, satisfying the usual $[x,p_x]=[y,p_y]=i\1$, with all the other commutators being
zero, to the capital operators, for which, again, $[X,P_X]=[Y,P_Y]=i\1$. However, $[X,Y]=i\theta\1$, while $[P_X,P_Y]=0$. These are the {\em non
commutative} operators we are going to consider here. They can be obtained from the original ones using, in a {\em non-standard fashion}, the unitary operator
$V_\theta:=e^{-i\frac{\theta}{2}p_xp_y}$: \be X=V_\theta x
V_\theta^{-1}, \quad P_X=V_\theta p_x V_\theta^{-1}, \quad Y=V_\theta^{-1} y V_\theta, \quad P_Y=V_\theta^{-1} p_y V_\theta. \label{22}\en
These show that each pair of variables transforms unitarily, but the two pairs transform in different ways. Thus  each single pair $(X,P_X)$ and $(Y,P_Y)$ is canonical, but the two pairs together are not.

One could try to introduce creation and annihilation operators in the usual way: $a_x=\frac{1}{\sqrt{2}}\left(x+ip_x\right)$ and
$a_y=\frac{1}{\sqrt{2}}\left(y+ip_y\right)$, and the related {\em capital} operators: $A_X=\frac{1}{\sqrt{2}}\left(X+iP_X\right)=V_\theta a_x
V_\theta^{-1}$ and $A_Y=\frac{1}{\sqrt{2}}\left(Y+iP_Y\right)=V_\theta^{-1} a_y V_\theta$. Since $[A_X,A_X^\dagger]=[A_Y,A_Y^\dagger]=\1$, one
could also think to construct eigenstates of $N_X=A_X^\dagger A_X$ and $N_Y=A_Y^\dagger A_Y$ in a standard fashion. This is possible, of course: one starts with, say,
the vacuum of $A_X$, $\Phi_0$,  $A_X \Phi_0=0$, and acts on this vector with powers of $A_X^\dagger$. Then
$N_X({A_X^\dagger}^n\Phi_0)=n({A_X^\dagger}^n\Phi_0)$. The same procedure can be repeated for $A_Y$. However, when we try to extend this
 construction to the two dimensional system, problems arise already at a very essential level: in fact, whenever the non commutativity
parameter $\theta$ is not zero, it is not possible to find any non zero, square-integrable, function $f(x,y)$ satisfying simultaneously
$A_Xf=0$ and $A_Yf=0$. In fact, using the definition of $A_X$ and $A_Y$, and the definitions in (\ref{21}), these conditions read as
$$
\left\{
    \begin{array}{ll}
        \left(x+\partial_x+i\frac{\theta}{2}\partial_y\right)f(x,y)=0  \\
        \left(y+\partial_y-i\frac{\theta}{2}\partial_x\right)f(x,y)=0.\\
       \end{array}
        \right.
$$
It is not hard to check that these equations cannot be solved together, except when $\theta=0$. In this case we recover the standard gaussian,
$f(x,y)=N\,e^{-(x^2+y^2)/2}$, which coincides with the common vacuum $\Phi_{0,0}(x,y)$ for $a_x$ and $a_y$, with the choice of the
normalization $N=\frac{1}{\sqrt{2\pi}}$. The conclusion is that, because of (\ref{21}), we lose the possibility of having a common vacuum for
the two annihilation operators $A_X$ and $A_Y$\footnote{We want to remark that this is not an automatic consequence of the fact that $[A_X,A_Y]\neq0$, since we are not trying to find a common set of eigenstates for these two operators, but just a single common vacuum.}.  Other more interesting choices of non commuting operators will be considered below.

\subsection{A different choice}

We have shown that the choice of the non commuting variables in (\ref{21}) is not appropriate if we are interested, for
instance, in extending the construction of the eigenstates of the hamiltonian of the Landau levels, since this construction is mainly based on
the existence of a common vacuum for two different annihilation operators. For this reason we discuss here a slightly different point of view,
in which the existence of such a vector is ensured by the construction itself. For this reason, rather than introducing {\em non commutative
coordinates}, it is more convenient to introduce directly {\em non commutative bosonic operators}.

Let $a_x$ and $a_y$ be as above, and let us introduce two linear combinations of these, \be A_1:=\alpha_x\,a_x+\alpha_y\,a_y,\qquad
A_2:=\beta_x\,a_x+\beta_y\,a_y, \label{2a}\en for some complex $\alpha$'s and $\beta$'s. This is the simplest non trivial choice we can
consider for our task. In fact, since $\Phi_{0,0}(x,y)$ is annihilated by both $a_x$ and $a_y$, it is also annihilated by $A_1$ and $A_2$, for all
choices of the coefficients: $A_1\Phi_{0,0}=A_2\Phi_{0,0}=0$. If we assume the following \be
|\alpha_x|^2+|\alpha_y|^2=|\beta_x|^2+|\beta_y|^2=1,\qquad \alpha_x\overline{\beta_x}+\alpha_y\overline{\beta_y}=\gamma, \label{2b}\en for some
$\gamma\in {\Bbb C}$, the following commutator rules follow: \be
        [A_1,A_2]=0,\quad [A_1,A_1^\dagger]=[A_2,A_2^\dagger]=\1,  \quad [A_1,A_2^\dagger]=\gamma\1
\label{23}\en which also imply that $[A_1^\dagger,A_2^\dagger]=0$ and that $[A_2,A_1^\dagger]=\overline{\gamma}\,\1$. In order to be able to invert
(\ref{2a}), we  also impose that \be \alpha_x\beta_y-\alpha_y\beta_x\neq0. \label{2c}\en This will be useful in a moment\footnote{It has been shown in \cite{balshali} that it is in fact possible to work with  more general $GL(n,\mathbb C)$ transformations to define the operators $A_i$, at least for the aspects concerning construction of the biorthogonal bases.}. Let us now proceed as usual: we define
\be
\Phi_{n_1,n_2}:=\frac{1}{\sqrt{n_1!\,n_2!}}(A_1^\dagger)^{n_1}(A_2^\dagger)^{n_2}\Phi_{0,0},
\label{23bis}\en
for $n_j\geq0$. Because of (\ref{2c}) it is possible to check that these vectors are linearly independent. This follows from the following fact:
let us construct the  vectors $\varphi_{n_1,n_2}:=\frac{1}{\sqrt{n_1!\,n_2!}}(a_x^\dagger)^{n_1}(a_y^\dagger)^{n_2}\Phi_{0,0}$. The set
$\F^\varphi:=\{\varphi_{n_1,n_2}, n_j\geq0\}$ is an orthonormal (o.n.) basis of $\Hil$. It is obvious that the
linear span of $\F^\Phi_M:=\{\Phi_{n_1,n_2},\,n_1+n_2=M\}$ coincides with that of $\{\varphi_{n_1,n_2},\,n_1+n_2=M\}$, for each $M=0,1,2,\ldots$. Therefore both these
sets span the same finite dimensional Hilbert space, $\Hil_M$, whose dimension is  $M+1$. Let us now introduce the operators
$N_j=A_j^\dagger A_j$, $j=1,2$. Due to the {\em unusual} commutation rules, $N_j$ is not a number operator. Indeed, using (\ref{23}), we find
$$
N_1\Phi_{n_1,n_2}=n_1\Phi_{n_1,n_2}+\gamma\sqrt{(n_1+1)n_2}\Phi_{n_1+1,n_2-1}, \quad N_2\Phi_{n_1,n_2}=n_2\Phi_{n_1,n_2}+\overline{\gamma}\sqrt{n_1(n_2+1)}\Phi_{n_1-1,n_2+1}.
$$
We see that two extra terms appear. However, two manifestly non self-adjoint number operators can be still introduced into the game. Indeed,
let us assume that $|\gamma|\neq1$. Then, if we introduce \be M_1=\frac{1}{1-|\gamma|^2}\left(N_1-\gamma A_1^\dagger \,A_2\right), \quad
M_2=\frac{1}{1-|\gamma|^2}\left(N_2-\overline{\gamma} A_2^\dagger \,A_1\right), \label{24}\en we get \be M_1\Phi_{n_1,n_2}=n_1\Phi_{n_1,n_2},
\quad M_2\Phi_{n_1,n_2}=n_2\Phi_{n_1,n_2}. \label{25}\en These two operators commute: $[M_1,M_2]=0$, so that it is not surprising they admit a
common set of eigenstates. However, since $M_j^\dagger\neq M_j$, we do not expect, in principle, that the $\Phi_{n_1,n_2}$ are mutually
orthogonal. Nevertheless, it is easy to see that some orthogonality can be established. For that we introduce the following operator: \be
H=M_1+M_2=\frac{1}{1-|\gamma|^2}\left(N_1+N_2-\gamma A_1^\dagger \,A_2-\overline{\gamma} A_2^\dagger \,A_1\right), \label{26}\en which is
clearly self-adjoint. Moreover $H\Phi_{n_1,n_2}=(n_1+n_2)\Phi_{n_1,n_2}$. This means the following: whenever $n_1+n_2\neq k_1+k_2$,
$$
\left\langle \Phi_{n_1,n_2},\Phi_{k_1,k_2}\right\rangle =0.
$$
In other words, taking $f\in \Hil_M$ and $g\in\Hil_L$, with $M\neq L$, then $<f,g>=0$. This is in agreement with what we have stated before,
i.e. that the linear span of $\{\Phi_{n_1,n_2},\,n_1+n_2=M\}$ coincides with that of $\{\varphi_{n_1,n_2},\,n_1+n_2=M\}$, and with the fact
that the different $\varphi_{n_1,n_2}$'s are mutually orthogonal.

Moreover, vectors $\Phi_{n_1,n_2}$ with the same value of $n_1+n_2$ are not, in general, mutually orthogonal. For instance, assuming that
$\|\Phi_{0,0}\|=1$,  we have \be \left\langle \Phi_{1,0},\Phi_{0,1}\right\rangle =\left\langle
A_1^\dagger\Phi_{0,0},A_2^\dagger\Phi_{0,0}\right\rangle = \left\langle \Phi_{0,0},\left([A_1,A_2^\dagger]+A_2^\dagger\,A_1
\right)\Phi_{0,0}\right\rangle =\gamma. \label{2bis}\en Analogously, we easily deduce that $\left\langle \Phi_{2,0},\Phi_{0,2}\right\rangle
=\gamma^2$, while $\left\langle \Phi_{2,0},\Phi_{1,1}\right\rangle =\left\langle \Phi_{1,1},\Phi_{0,2}\right\rangle =\sqrt{2}\,\gamma$, and so
on.

Being linearly independent, however, the $M+1$ vectors of $\F_M^\Phi$ form a basis of the $M+1$-dimensional space $\Hil_M$. In what follows we
will discuss how, extending what was done in \cite{bagpf}, it is possible to construct a second family of vectors which is biorthogonal to
$\F_M^\Phi$. These vectors are expected, among other things, to be eigenvectors of $M_1^\dagger$ and $M_2^\dagger$. We will return on this
aspect later on.

Since the role of finite-dimensional Hilbert spaces will be quite important in what follows, we want to stress that finite-dimensional systems are, quite often, very important in quantum mechanics, and in non hermitian quantum mechanics as well, since they usually allow a reasonably simple comprehension of some aspects of the theory which, otherwise, would be hidden by the many (mathematical) technicalities which are intrinsic to infinite-dimensional systems. The recent literature on this subject is rather rich and a complete list would be too long. We just want to cite here some of those papers which are more relevant for us, \cite{fdmod}.

\subsection{Relation with extended pseudo-fermions}

It is now convenient to rename the vectors of $\Hil_M$ as follows
$$
h_0^{(M)}=\Phi_{M,0},\quad h_1^{(M)}=\Phi_{M-1,1},\ldots,\quad h_M^{(M)}=\Phi_{0,M},
$$
and let $\F_M^{(h)}$ be the set of these $M+1$ vectors: $\F^{(h)}_M=\{h_0^{(M)}, h_1^{(M)},\ldots,h_M^{(M)}\}$. Of course, $\F_M^{(h)}$
coincides with the set $\F_M^\Phi$ introduced previously. For each fixed $M$ these vectors are linearly independent for those choice of
$\alpha$'s and $\beta$'s which satisfy (\ref{2b}) and (\ref{2c}). Moreover, vectors $h_i^{(M)}$ and $h_j^{(M')}$ are automatically orthogonal if $M\neq M'$.

Following and extending the original idea introduced in \cite{bagpf} we will show now how a rather general algebraic procedure, again related to
some suitable deformation of the anticommutation relations and in particular to some generalized raising and lowering operators, can be introduced into the game in order to produce, in each finite-dimensional
Hilbert space $\Hil_M$, a new family of vectors,  $\F^{(e)}_M=\{e_0^{(M)}, e_1^{(M)},\ldots,e_M^{(M)}\}$, which is biorthogonal to the vectors
in $\F^{(h)}_M$ (see also, \cite{balshali}) and how these can be used to construct some families of intertwining operators. For that, after considering briefly what
happens in $\Hil_0$, we will show in detail how the procedure works in $\Hil_1$ and $\Hil_2$, and then we will discuss how to extend this procedure to
higher values of $M$. Needless to say, we are not claiming here that ours is the only procedure which produces two biorthogonal families in a finite-dimensional Hilbert space, or that this is crucially related to non commutative quantum mechanics: what we are saying is that the procedure we are going to describe is natural and interesting both from a mathematical side and for possible physical applications.

\vspace{2mm}

First of all, it is obvious that the  set $\F^{(e)}_0$ simply coincides with $\F^{(h)}_0$, except possibly for a normalization factor: indeed,
if we define $e_0^{(0)}:=\frac{1}{\|h_0^{(0)}\|^2}\,h_0^{(0)}$, then $\left\langle e_0^{(0)},h_0^{(0)}\right\rangle =1$. Both sets are bases in the 1-dimensional Hilbert space $\Hil_0$.

More interesting is the situation for $M=1$. In this case we introduce the bounded operators $a_1$ and $b_1$ via their action on the basis
$\F^{(h)}_1$: \be a_1\,h_0^{(1)}:=0,\quad a_1\,h_1^{(1)}:=h_0^{(1)}, \quad\mbox{and}\quad b_1\,h_0^{(1)}:=h_1^{(1)},\quad b_1\,h_1^{(1)}:=0.
\label{31}\en We see that $a_1$ and $b_1$ act as lowering and raising operators on $\F^{(h)}_1$. From this definition we deduce that \be
a_1^2=0\quad b_1^2=0,\quad \{a_1,b_1\}=\1_1, \label{32}\en where $\1_1$ is the identity operator on $\Hil_1$. These are exactly the
pseudo-fermionic anti-commutation rules considered in \cite{bagpf}, so that the same construction proposed there can be repeated here. The
starting point is a non-zero vector, $e_0^{(1)}$, orthogonal to $h_1^{(1)}$. Such a vector surely exists, since $\mathrm{dim}(\Hil_1)=2$.
Moreover, it is always possible to choose its normalization in such a way $\left\langle e_0^{(1)},h_0^{(1)}\right\rangle =1$. It is easy to
check that $e_0^{(1)}$ is the vacuum for $b_1^\dagger$: $b_1^\dagger\,e_0^{(1)}=0$. In fact, taken a generic vector $f\in\Hil_1$ and recalling
that $f$ can be written as $f=c_0\,h_0^{(1)}+c_1\,h_1^{(1)}$, for some complex $c_0$ and $c_1$, using (\ref{31}) we deduce that $\left\langle
f,b_1^\dagger \,e_0^{(1)}\right\rangle =\overline{c_0} \left\langle h_1^{(1)},e_0^{(1)}\right\rangle =0$. Then our claim follows from the
arbitrariness of $f$.

Let us now define the vector $e_1^{(1)}:=a_1^\dagger e_0^{(1)}$. Since
$\left\langle e_1^{(1)},h_0^{(1)}\right\rangle =\left\langle e_0^{(1)},a_1\,h_0^{(1)}\right\rangle =0$ and
$\left\langle e_1^{(1)},h_1^{(1)}\right\rangle =\left\langle e_0^{(1)},a_1\,h_1^{(1)}\right\rangle =\left\langle e_0^{(1)},h_0^{(1)}\right\rangle =1$, we conclude that
$\F^{(e)}_1=\{e_0^{(1)}, e_1^{(1)}\}$ is biorthonormal to $\F^{(h)}_1$. These two sets are respectively eigenstates of
$N_1^\dagger=a_1^\dagger\,b_1^\dagger$ and $N_1=b_1\,a_1$:
$$
N_1\,h_k^{(1)}=k\,h_k^{(1)},\qquad N_1^\dagger\,e_k^{(1)}=k\,e_k^{(1)},
$$
for $k=0,1$. Moreover, they resolve the identity in $\Hil_1$:
$$
\sum_{k=0}^1\,|e_k^{(1)}\left>\right<h_k^{(1)}|=\sum_{k=0}^1\,|h_k^{(1)}\left>\right<e_k^{(1)}|=\1_1.
$$
We can also introduce two self-adjoint, positive and invertible operators
$$
S_1^{(h)}=\sum_{k=0}^1\,|h_k^{(1)}\left>\right<h_k^{(1)}|,\qquad S_1^{(e)}=\sum_{k=0}^1\,|e_k^{(1)}\left>\right<e_k^{(1)}|,
$$
or, more explicitly,
$$
S_1^{(h)}\,f=\sum_{k=0}^1\,\left\langle h_k^{(1)},f\right\rangle \,h_k^{(1)},\qquad S_1^{(e)}\,f=\sum_{k=0}^1\,\left\langle e_k^{(1)},f\right\rangle \,e_k^{(1)},
$$
for each $f\in\Hil_1$. These operators are inverses of one another: $S_1^{(h)}=\left(S_1^{(e)}\right)^{-1}$. Moreover, they map $\F^{(e)}_1$
into $\F^{(h)}_1$ and viceversa:
$$
S_1^{(h)}\,e_k^{(1)}=h_k^{(1)},\quad S_1^{(e)}\,h_k^{(1)}=e_k^{(1)},
$$
$k=0,1$, and they satisfy the following intertwining relations:
$$
S_1^{(e)}\,N_1=N_1^\dagger\, S_1^{(e)},\qquad N_1\, S_1^{(h)}=S_1^{(h)}\,N_1^\dagger.
$$
There is something more: since they are positive operators, the square roots of $S_1^{(h)}$ and $S_1^{(e)}$ surely exist. Therefore, we can define
$$
n_1:=\left(S_1^{(e)}\right)^{1/2}\,N_1\,\left(S_1^{(h)}\right)^{1/2},\qquad c_k^{(1)}:=\left(S_1^{(e)}\right)^{1/2}\, h_k^{(1)},
$$
$k=0,1$. It is easy to check that $n_1$ is a self-adjoint operator on $\Hil_1$, and that $\F^{(c)}_1=\{c_0^{(1)}, c_1^{(1)}\}$ is an o.n. basis
of $\Hil_1$.

\vspace{3mm}

A similar procedure can be repeated also for $\Hil_2$. In this case, however, we lose the relations in (\ref{32}), but we still
maintain the main aspects of the functional structure. The starting point is, as before, the basis $\F^{(h)}_2=\{h_0^{(2)}, h_1^{(2)},h_2^{(2)}\}$. In this
case the raising and lowering operators, $b_2$ and $a_2$, are defined by an extended version of (\ref{31}):
\be
a_2\,h_0^{(2)}:=0,\quad a_2\,h_1^{(2)}:=h_0^{(2)},\quad a_2\,h_2^{(2)}:=\sqrt{2}\,h_1^{(2)},
\label{3a1}\en
and
\be
b_2\,h_0^{(2)}:=h_1^{(2)},\quad b_2\,h_1^{(2)}:=\sqrt{2}\,h_2^{(2)},\quad b_2\,h_2^{(2)}:=0.
\label{3b1}\en
In this case $a_2^3=b_2^3=0$, but $\{a_2,b_2\}\neq\1_2$. Nevertheless, if we define $N_2=b_2\,a_2$, we still get $N_2\,h_k^{(2)}=k\,h_k^{(2)}$,
$k=0,1,2$, so that the vectors $h_k^{(2)}$ are eigenstates of a number-like operator. The biorthogonal set $\F^{(e)}_2$ is now constructed
extending the previous procedure: we begin considering a vector, $e_0^{(2)}$, which is orthogonal to both  $h_1^{(2)}$ and $h_2^{(2)}$. This
vector is unique up to a normalization, which we choose  in such a way that $\left\langle e_0^{(2)},h_0^{(2)}\right\rangle =1$. We find that
$b_2^\dagger\,e_0^{(2)}=0$. Defining further $e_1^{(2)}:=a_2^\dagger\,e_0^{(2)}$ and $e_2^{(2)}:=\frac{1}{\sqrt{2}}\,a_2^\dagger\,e_1^{(2)}$,
we get
$$
\left\langle e_j^{(2)},h_k^{(2)}\right\rangle =\delta_{j,k},
$$
$j,k=0,1,2$. Hence $\F^{(e)}_2$ into $\F^{(h)}_2$ are biorthogonal bases of $\Hil_2$. The vector $e_k^{(2)}$ is eigenstate of $N_2^\dagger$:
$N_2^\dagger\,e_k^{(2)}=k\,e_k^{(2)}$, $k=0,1,2$. This can be proved by using the lowering nature of $b_2^\dagger$ on $\F^{(e)}_2$. In fact,
other than $b_2^\dagger\,e_0^{(2)}=0$, we can also check that $b_2^\dagger\,e_1^{(2)}=e_0^{(2)}$, and that
$b_2^\dagger\,e_2^{(2)}=\sqrt{2}\,e_1^{(2)}$. Two operators, $S_2^{(h)}$ and $S_2^{(e)}$, can be defined as before, and for these we can prove
exactly analogous results as those for $S_1^{(h)}$ and $S_1^{(e)}$. Hence, what appears to be really relevant in this construction, is not
really the anticommutation rule $\{a,b\}=\1$, but the definition of the raising and lowering operators. For this reason, we call these {
particles}, {\em generalized pseudo-fermions}. Other interesting generalizations of fermions could be found, for instance, in \cite{tri1,tri2}
and in \cite{ortoferm}.

For generic $M$ we could repeat the same construction, starting from $\F^{(h)}_M$. The two operators $a_M$ and $b_M$, defined extending formulas (\ref{3a1}) and (\ref{3b1}), satisfy the following
property: $a_M^{M+1}=b_M^{M+1}=0$. As for the anti-commutator rule, we can write
$\{a_M,b_M\}=\sum_{k=0}^M\alpha_k^{(M)}\,|e_k^{(M)}\left>\right<h_k^{(M)}|$, where the coefficients can be easily found. For instance we have
$\alpha_0^{(1)}=\alpha_1^{(1)}=1$, $\alpha_0^{(2)}=1$, $\alpha_1^{(2)}=3$ and $\alpha_2^{(2)}=2$, and yet $\alpha_0^{(3)}=1$,
$\alpha_1^{(3)}=3$, $\alpha_2^{(3)}=5$ and $\alpha_3^{(3)}=3$. In matrix form we have:
$$\{a_M,b_M\}=\left(
                \begin{array}{cccccccc}
                  1 & 0 & 0 & 0 & \cdot & \cdot & 0 & 0 \\
                  0 & 3 & 0 & 0 & \cdot & \cdot & 0 & 0 \\
                  0 & 0 & 5 & 0 & \cdot & \cdot & 0 & 0 \\
                  0 & 0 & 0 & 7 & \cdot & \cdot & 0 & 0 \\
                  0 & 0 & 0 & 0 & \cdot & \cdot & 0 & 0 \\
                  0 & 0 & 0 & 0 & \cdot & \cdot & 0 & 0 \\
                  0 & 0 & 0 & 0 & \cdot & \cdot & 2M-1 & 0 \\
                  0 & 0 & 0 & 0 & \cdot & \cdot & 0 & M \\
                \end{array}
              \right),
$$
which, for instance, taking $M=1$, gives back $\{a_M,b_M\}=\1_1$: $M=1$ is the only choice which furnishes the identity in the right-hand side of $\{a_M,b_M\}$. The vectors of $\F^{(h)}_M$ are eigenstates of $N_M=b_Ma_M$, while those of the biorthogonal set
$\F^{(e)}_M=\left\{e_l^{(M)},\,l=0,1,\ldots,M\right\}$, are eigenstates of its adjoint
$N_M^\dagger$, and we have $\left\langle e_l^{(M)},h_k^{(N)}\right\rangle =\delta_{l,k}\delta_{M,N}$.

\section{An explicit realization}

It is interesting now to see how the above operators and bases can be explicitly constructed.  In particular, we will show that the main ingredient of the construction is provided by the overlaps between the different vectors in $\F_M^{(h)}$: these are, in turn, fixed by the commutation rules in (\ref{23}) and by the definition of the vectors $\Phi_{n_1,n_2}$, (\ref{23bis}), as briefly shown, for instance, by formula (\ref{2bis}): this is the way in which the non commutativity of the bosons came into the game. Once these coefficients are found, there is not a
single way to chose the $M+1$-dimensional vectors $h_k^{(M)}$ which reproduce these overlaps. Different realizations, i.e. different choices of $\F_M^{(h)}$, are possible. However, once this first set is chosen, the biorthogonal set $\F_M^{(e)}$ is uniquely determined. Stated differently, we are changing here our point of view, focusing on the  values of $\left\langle \Phi_{n_1,n_2},\Phi_{k_1,k_2}\right\rangle$, for $n_1+n_2=k_1+k_2$, rather than on the vectors $\Phi_{m_1,m_2}$ themselves. Once these values are found, we will look for those finite-dimensional vectors $h_k^{(M)}$ which produce, inside each $\Hil_M$, these particular overlaps. In practice, we will now represent each infinite-dimensional vector $\Phi_{m_1,m_2}$ as a finite dimensional vector in $\Hil_{m_1+m_2}$.

Of course,  $\Hil_0$ being a one-dimensional space, there is not much to say. In this case $h_0^{(0)}=e_0^{(0)}$, with a normalization chosen in
such a way $\left\langle h_0^{(0)},e_0^{(0)}\right\rangle =1$, and $a_0$, $b_0$ both annihilate these states.

\subsection{$M=1$}

Here the situation is more interesting. To avoid useless complications, from now on we consider $\gamma>0$. As we have discussed above, in
order to produce our biorthogonal set and the related operators,  we just need
 to find two two-dimensional vectors $h_j^{(1)}$ which represent $\Phi_{k,l}$, $j=0,1$ and $k+l=1$: we simply write $h_0^{(1)}=\Phi_{1,0}$ and $h_1^{(1)}=\Phi_{0,1}$, and we ask for $\left\langle h_0^{(1)},h_1^{(1)}\right\rangle =\gamma$, see (\ref{2bis}). A possible choice
(clearly highly non-unique!) of these vectors is the following:
$$
h_0^{(1)}=\sqrt{\gamma}\,\left(
                           \begin{array}{c}
                             1 \\
                             0 \\
                           \end{array}
                         \right), \qquad h_1^{(1)}=\sqrt{\gamma}\,\left(
                           \begin{array}{c}
                             1 \\
                             1 \\
                           \end{array}
                         \right).
$$
Then, the two two-by-two matrices $a_1$ and $b_1$ which satisfy the raising and lowering identities given in (\ref{31}) are the following:
$$
a_1=\left(
      \begin{array}{cc}
        0 & 1 \\
        0 & 0 \\
      \end{array}
    \right),\qquad b_1=\left(
      \begin{array}{cc}
        1 & -1 \\
        1 & -1 \\
      \end{array}
    \right).
$$
It is clear that $a_1^2=b_1^2=0$, and that $\{a_1,b_1\}=\1_1$. Moreover $N_1=b_1a_1=\left(
      \begin{array}{cc}
        0 & 1 \\
        0 & 1 \\
      \end{array}
    \right)$, which is not self-adjoint. But $N_1$ is still a number-like operator, since $N_1\,h_k^{(1)}=k\,h_k^{(1)}$, $k=0,1$. Let us now
    construct the biorthonormal basis $\F_1^{(e)}$. The first step consists in finding the kernel of $b_1^\dagger$.
    This is one dimensional and it is proportional to the vector $\left(
                           \begin{array}{c}
                             1 \\
                             -1 \\
                           \end{array}
                         \right)$. Hence, in order to have $\left\langle h_0^{(1)},e_0^{(1)}\right\rangle =1$, we take $e_0^{(1)}=\frac{1}{\sqrt{\gamma}}\,\left(
                           \begin{array}{c}
                             1 \\
                             -1 \\
                           \end{array}
                         \right)$. Then $e_1^{(1)}=a_1^\dagger\,e_0^{(1)}=\frac{1}{\sqrt{\gamma}}\,\left(
                           \begin{array}{c}
                             0 \\
                             1 \\
                           \end{array}
                         \right)$. It is now an easy exercise to check the resolutions of the identity
                         $\sum_{k=0}^1|h_k^{(1)}\left>\right<e_k^{(1)}| = \sum_{k=0}^1|e_k^{(1)}\left>\right<h_k^{(1)}| = \1_1$,
                         and to deduce the matrix form of the intertwining operators
$$
S_1^{(h)}=\sum_{k=0}^1|h_k^{(1)}\left>\right<h_k^{(1)}|=\gamma\left(
                                                          \begin{array}{cc}
                                                            2 & 1 \\
                                                            1 & 1 \\
                                                          \end{array}
                                                        \right),
$$
and
$$
S_1^{(e)}=\sum_{k=0}^1|e_k^{(1)}\left>\right<e_k^{(1)}|=\frac{1}{\gamma}\left(
                                                          \begin{array}{cc}
                                                            1 & -1 \\
                                                            -1 & 2 \\
                                                          \end{array}
                                                        \right).
$$
For these matrices our formulas can be easily checked. For instance they are both self-adjoint, one is the inverse of the other, and
$S_1^{(h)}\,e_k^{(1)}=h_k^{(1)}$, $k=0,1$. Also, $S_1^{(e)}\,N_1=N_1^\dagger\,S_1^{(e)}$. Taking {one of the square roots of $S_1^{(e)}$ as},
$$
\left(S_1^{(e)}\right)^{1/2}=\frac{1}{\sqrt{5\,\gamma}}\left(
                                                          \begin{array}{cc}
                                                            2 & -1 \\
                                                            -1 & 3 \\
                                                          \end{array}
                                                        \right),
$$
and its inverse, it is possible to define a new self-adjoint number operator,
$$n_1=\left(S_1^{(e)}\right)^{1/2}\,N_1\,\left(S_1^{(e)}\right)^{-1/2}=\frac{1}{5}\left(
                                                          \begin{array}{cc}
                                                            1 & 2 \\
                                                            2 & 4 \\
                                                          \end{array}
                                                        \right),$$ and its eigenvectors $c_0^{(1)}=\left(S_1^{(e)}\right)^{1/2}\,h_0^{(1)}=\frac{1}{\sqrt{5}}\left(
                           \begin{array}{c}
                             2 \\
                             -1 \\
                           \end{array}
                         \right)$, and $c_1^{(1)}=\left(S_1^{(e)}\right)^{1/2}\,h_1^{(1)}=\frac{1}{\sqrt{5}}\left(
                           \begin{array}{c}
                             1 \\
                             2 \\
                           \end{array}
                         \right)$, which, as we can see, is an o.n. basis in $\Hil_1$. Incidentally we observe that these vectors are different from the
                         canonical basis in ${\Bbb C}^2$.

\subsection{$M=2$}

Similar computations can be performed also for $\Hil_2$. Again, the starting point is a fixed choice of three linearly independent vectors $h_j^{(2)}$ which represent $\Phi_{k,l}$, $j=0,1,2$ and $k+l=2$: again we simply write
$h_0^{(2)}=\Phi_{2,0}$, $h_1^{(2)}=\Phi_{1,1}$ and $h_2^{(2)}=\Phi_{0,2}$, and we ask that $\left\langle h_0^{(2)},h_1^{(2)}\right\rangle
=\left\langle h_1^{(2)},h_2^{(2)}\right\rangle =\gamma\sqrt{2}$ and $\left\langle h_0^{(2)},h_2^{(2)}\right\rangle =\gamma^2$, see Section II.
Again, this choice is not unique. Here we consider the following:
$$
h_0^{(2)}=\gamma\sqrt{2}\left(
                          \begin{array}{c}
                            1 \\
                            0 \\
                            0 \\
                          \end{array}
                        \right),\qquad h_1^{(2)}=\left(
                          \begin{array}{c}
                            1 \\
                            \gamma/\sqrt{2} \\
                            0 \\
                          \end{array}
                        \right),\qquad h_2^{(2)}=\left(
                          \begin{array}{c}
                            \gamma/\sqrt{2} \\
                            1 \\
                            1 \\
                          \end{array}
                        \right).
$$
Hence the matrices $a_2$ and $b_2$ are uniquely found to be
$$
a_2=\left(
      \begin{array}{ccc}
        0 & 2 & \sqrt{2}-2 \\
        0 & 0 & \gamma \\
        0 & 0 & 0 \\
      \end{array}
    \right),\qquad b_2=\left(
      \begin{array}{ccc}
        \frac{1}{\sqrt{2}\,\gamma} & \frac{\sqrt{2}}{\gamma}\left(\gamma-\frac{1}{\sqrt{2}\,\gamma}\right) & -\frac{1}{2}-\sqrt{2}+\frac{1}{\gamma^2} \\
        \frac{1}{2} & \frac{\sqrt{2}}{\gamma}\left(\sqrt{2}-\frac{1}{2}\right) & -\frac{\gamma}{2\sqrt{2}}-\frac{\sqrt{2}}{\gamma}\left(\sqrt{2}-\frac{1}{2}\right) \\
        0 & \frac{2}{\gamma} & -\frac{2}{\gamma} \\
      \end{array}
    \right).
$$
They satisfy, in particular, the condition $a_2^3=b_2^3=0$. The form of the (non self-adjoint) number operator is now
$$
N_2=b_2a_2=\left(
      \begin{array}{ccc}
        0 & \frac{\sqrt{2}}{\gamma} & \frac{\sqrt{2}(\gamma^2-1)}{\gamma} \\
        0 & 1 & 1 \\
        0 & 0 & 2 \\
      \end{array}
    \right).
$$
Then $N_2h_k^{(2)}=k\,h_k^{(2)}$. In order to fix the vectors  $e_k^{(2)}$ we start again looking for the kernel of $b_2^\dagger$. The
normalization is fixed by requiring that  $\left\langle e_0^{(2)},h_0^{(2)}\right\rangle =1$. Then we define $e_1^{(2)}=a_2^\dagger e_0^{(2)}$ and
$e_2^{(2)}=\frac{1}{\sqrt{2}}\,a_2^\dagger e_1^{(2)}$:
$$
e_0^{(2)}=\frac{1}{\gamma^2}\left(
                          \begin{array}{c}
                            \frac{\gamma}{\sqrt{2}} \\
                            -1 \\
                            1-\frac{\gamma^2}{2} \\
                          \end{array}
                        \right),\qquad e_1^{(2)}=\frac{\sqrt{2}}{\gamma}\left(
                          \begin{array}{c}
                            0 \\
                            1 \\
                            -1 \\
                          \end{array}
                        \right),\qquad e_2^{(2)}=\left(
                          \begin{array}{c}
                            0 \\
                            0 \\
                            1 \\
                          \end{array}
                        \right).
$$
Hence $\left\langle e_k^{(2)},h_n^{(2)}\right\rangle =\delta_{k,n}$. It is easy to recover the resolution of the identity in $\Hil_2$, and to deduce the
expressions for the  operators $S_2^{(h)}$ and  $S_2^{(e)}$:
$$
S_2^{(h)}=\left(
      \begin{array}{ccc}
        1+\frac{5}{2}\gamma^2 & \sqrt{2}\,\gamma & \frac{\gamma}{\sqrt{2}} \\
        \sqrt{2}\,\gamma & 1+\frac{\gamma^2}{2} & 1 \\
        \frac{\gamma}{\sqrt{2}} & 1 & 1 \\
      \end{array}
    \right),
$$
and
$$
S_2^{(e)}=\left(
      \begin{array}{ccc}
        \frac{1}{2\gamma^2} & -\frac{1}{\sqrt{2}\,\gamma^3} & \frac{1}{\sqrt{2}\,\gamma^3}\left(1-\frac{\gamma^2}{2}\right) \\
        -\frac{1}{\sqrt{2}\,\gamma^3} & \frac{1}{\gamma^2}\left(2+\frac{1}{\gamma^2}\right) & -\frac{1}{\gamma^2}\left(\frac{3}{2}+\frac{1}{\gamma^2}\right) \\
        \frac{1}{\sqrt{2}\,\gamma^3}\left(1-\frac{\gamma^2}{2}\right) & -\frac{1}{\gamma^2}\left(\frac{3}{2}+\frac{1}{\gamma^2}\right) & \frac{1}{\gamma^4}+\frac{1}{\gamma^2}+\frac{5}{4} \\
      \end{array}
    \right).
$$
Even for these matrices it is possible to check all the properties found previously. For instance, they are one the inverse of the other, and
they intertwine between $N_2$ and $N_2^\dagger$. The analytic expression for {(one of) their square roots} is more complicated, and will not be given here.

\vspace{3mm}

Extending what we have discussed so far, we can  conclude that {\em in any finite dimensional Hilbert space $\hat\Hil$, given a non-orthogonal basis, it is
possible introduce a pair of generalized pseudo-fermionic operators $a$ and $b$ which give rise, looking at the kernel of $b^\dagger$ and
acting on this vector with powers of $a^\dagger$ (with suitable normalization factors), to another basis of $\hat\Hil$ which is biorthonormal
to the first one. Both bases are eigenstates of two non self-adjoint number-like operators. Moreover, interesting intertwining relations between
these two operators can be deduced.}

\vspace{2mm}

It might be worth noticing that, while the existence and the construction of  biorthogonal bases are widely discussed in the mathematical literature, the procedure described here has many aspects which, in our knowledge, have not been considered before. In particular, relations with self-adjoint or crypto-hermitian number-like operators and the existence of some intertwining relations are, in our opinion, peculiar of the present construction.

\section{Back to $\Hil$}

In each $\Hil_M$, due to the fact that finite matrices are always bounded operators, it is clear that each set $\F_M^{(h)}$ or $\F_M^{(e)}$ are
Riesz bases. This is true for each fixed value of $M$, but appears not to be necessarily so when we go back to $\Hil$, since
$\Hil=\bigoplus_{M=0}^\infty\Hil_M$. Therefore, even if it is reasonable to expect that $\F^{(h)}=\{h_k^{(M)}, M\geq0, \, k=0,1,2,\ldots,M\}$
and $\F^{(e)}=\{e_k^{(M)}, M\geq0, \, k=0,1,2,\ldots,M\}$ are indeed (biorthogonal), but not necessarily Riesz, bases for $\Hil$, this remains
an open problem. However, supposing that this is indeed the case, using the mutual orthogonality between different $\Hil_M$, we can write
$$
\1=\sum_{M=0}^\infty\1_M=\sum_{M=0}^\infty\sum_{l=0}^M|e_l^{(M)}\left>\right<h_l^{(M)}|=\sum_{M=0}^\infty\,P_M,
$$
where $P_M:=\sum_{l=0}^M|e_l^{(M)}\left>\right<h_l^{(M)}|=\1_M$ is an orthogonal projection on $\Hil_M$: $P_M=P_M^2=P_M^\dagger$.  Moreover, if
$M\neq N$, $P_M\,P_N=0$. Now we define \be A=\sum_{M=0}^\infty P_M\,a_M\,P_M,\qquad B=\sum_{M=0}^\infty P_M\,b_M\,P_M. \label{41}\en These
operators are not defined on the whole $\Hil$, but on the domains
$$
D(A)=\left\{f\in\Hil: \left\{f^A_N:=\sum_{M=1}^N\sum_{l=1}^M\left\langle e_l^{(M)},f\right\rangle \sqrt{l}\,h_{l-1}^{(M)}\right\} \,\mbox{is a Cauchy sequence in } \Hil\right\},
$$
while
$$
D(B)=\left\{f\in\Hil: \left\{f^B_N:=\sum_{M=1}^N\sum_{l=0}^{M-1}\left\langle e_l^{(M)},f\right\rangle \sqrt{l+1}\,h_{l+1}^{(M)}\right\} \,\mbox{is a Cauchy sequence in } \Hil\right\}.
$$
Then, $\forall f\in D(A)$ and $\forall g\in D(B)$, we find
$$
Af=\sum_{M=1}^\infty\sum_{l=1}^M\left\langle e_l^{(M)},f\right\rangle \sqrt{l}\,h_{l-1}^{(M)}, \quad
Bg=\sum_{M=1}^\infty\sum_{l=0}^{M-1}\left\langle e_l^{(M)},g\right\rangle \sqrt{l+1}\,h_{l+1}^{(M)},
$$
which are convergent by construction. The above domains are dense in $\Hil$, since they both contain $\F^{(h)}$, which, as stated, is a basis
for $\Hil$. In particular, recalling that $k=0,1,2,\ldots,M$, we find that
$$
A\,h_k^{(M)}=\left\{
\begin{array}{ll}
0,\hspace{5cm} \mbox{if } k=0, \,\forall M\\
    \sqrt{k}\,h_{k-1}^{(M)}\hspace{3.9cm} \mbox{if } M\geq1, \,\forall k=1,2,\ldots,M,\\
     \end{array}
        \right.
$$
while
$$
B\,h_k^{(M)}=\left\{
\begin{array}{ll}
0,\hspace{5cm} \mbox{if } k=M, \,\forall M\\
    \sqrt{k+1}\,h_{k+1}^{(M)}\hspace{3.3cm} \mbox{otherwise. }\\
     \end{array}
        \right.
$$
We can now interpret the operator $N=BA$ as a (partial) number operator on $\Hil$, since $N\,h_k^{(M)}=k\,h_k^{(M)}$ for all $k$ and $M$. Of
course, $D(N)$ is the following proper subspace of $\Hil$:  $D(N)=\{f\in D(A):\,Af\in D(B)\}$. The operators $A^\dagger$ and $B^\dagger$ are
densely defined, since $D(A^\dagger)$ and $D(B^\dagger)$ both contain $\F^{(e)}$, see below, which is also a basis for $\Hil$. In particular,
for our present purposes, it is enough to notice that
$$
A^\dagger\,e_k^{(M)}=\left\{
\begin{array}{ll}
0,\hspace{3.5cm} \mbox{if } k=M, \,\forall M\\
    \sqrt{k+1}\,e_{k+1}^{(M)}\hspace{1.9cm} \mbox{if }  k=0,1,\ldots,M-1,\\
     \end{array}
        \right.
$$
while
$$
B^\dagger\,e_k^{(M)}=\left\{
\begin{array}{ll}
0,\hspace{4.3cm} \mbox{if } k=0, \,\forall M\\
    \sqrt{k}\,e_{k-1}^{(M)}\hspace{3.3cm} \mbox{otherwise. }\\
     \end{array}
        \right.
$$
In fact, let us first introduce $D(A^\dagger)=\{g\in\Hil \mbox{ such that } \exists g_A\in\Hil:\,
\left<g_A,f\right>=\left<g,A\,f\right>,\,\forall\,f\in D(A)\}$. Then, $\forall \,g\in D(A^\dagger)$, $A^\dagger$ is defined as
$A^\dagger\,g:=g_A$. Now, taken $f\in D(A)$, it is easy to check that $e_k^{(M)}\in D(A^\dagger)$ and that the first formula above holds. For
that we observe that, because of the above expansion for $Af$,
$$
\left<A^\dagger e_k^{(M)}, f\right>=\left<e_k^{(M)},A f\right>=\sum_{N=1}^\infty \sum_{l=1}^N\left\langle e_l^{(N)},f\right\rangle
\sqrt{l}\,\left\langle e_k^{(M)}, h_{l-1}^{(N)}\right\rangle=$$ $$=\sum_{N=1}^\infty \sum_{l=1}^N\left\langle e_l^{(N)},f\right\rangle
\sqrt{l}\,\delta_{N,M}\delta_{l-1,k}=\left\langle \sqrt{k+1}\,e_{k+1}^{(M)},f\right\rangle,
$$
where we have considered here $k$ ranging between 0 and $M-1$.  Our conclusion follows from the arbitrariness of $f$. The case $k=M$ is even simpler. With similar computations we could
also check the formula for $B^\dagger\,e_k^{(M)}$.

Now, calling $N^\sharp=A^\dagger B^\dagger$, we see that $N^\sharp e_k^{(M)}=k\,e_k^{(M)}$. Of course, $D(N^\sharp)$ is a proper subspace of
$\Hil$, defined in analogy with $D(N)$\footnote{We refer to \cite{bit} for all the mathematical subtleties which are related to the unbounded
nature of the operators introduced in this section, and which are not extremely important for us here, since we are essentially interested in
what happens in each $\Hil_M$.}. It is now natural to introduce two operators acting on $\F^{(e)}$ and $\F^{(h)}$ in the following way: \be
S^{(h)}\,e_k^{(M)}=h_k^{(M)}, \qquad S^{(e)}\,h_k^{(M)}=e_k^{(M)}, \label{42}\en and then extend these definitions to finite linear
combinations of these vectors. It is clear that, in each subspace $\Hil_M$, these operators coincide with those considered in the previous
sections:
$$
S^{(h)}\upharpoonright_{\Hil_M}=S^{(h)}_M,\qquad S^{(e)}\upharpoonright_{\Hil_M}=S^{(e)}_M.
$$
Hence they are bounded on each subspace. However, this does not imply that they are also {\em globally} bounded, i.e. bounded on $\Hil$.
Exactly for this reason, \cite{bagpb10}, the intertwining relation should also be considered with a certain care: in fact, condition
$S_M^{(e)}\,N_M=N_M^\dagger\,S_M^{(e)}$, for each fixed $M\in{\Bbb N}$, does not imply also that $S^{(e)}\,N=N^\dagger\,S^{(e)}$. We can only
conclude that $\left(S^{(e)}\,N-N^\dagger\,S^{(e)}\right)\,h_k^{(M)}=0$, for all $k$ and $M$. However, in the present situation, this is not
enough to recover the operatorial intertwining relation, \cite{hal}. This is due to the fact that the operators involved are unbounded. Notice
also that they can be represented as diagonal block matrices, the $M-th$ block being a matrix acting on $\Hil_M$. Each block represents a
bounded operator, but the infinite matrix is not necessarily bounded.

\section{Coherent states}

In the literature there does not exist  full agreement on what a (generalized) coherent state should be: for some authors some properties are more important than others, and this leads to very different generalizations, \cite{gazbook}. Here, we will concentrate on a particular aspect of these states, which is essential for quantization, and thus insist on a resolution of the identity, in an appropriate sense. This is the point of view of, say, \cite{gazetal}, where the authors are not so much interested in whether  these states are eigenstates of some lowering operator or not.

The idea is quite simple: let $X$ be some subset of ${\bf R}^d$, $d=1,2,3,\ldots$, equipped with a measure $\mu$ and let $\Lc^2(X,d\mu)$ the set of all the square-integrable Lebesgue-measurable functions on $X$, with scalar product
$$
\left\langle f,g\right\rangle _{\Lc^2}:=\int_X\overline{f(x)}\,g(x)d\mu(x).
$$
Let us select, in $\Lc^2(X,d\mu)$, a family of functions $\{\Phi_n(x), \,x\in X: \,n=0,1,\ldots,N-1\}$, where $N$ could be finite or not. We require that $\left\langle \Phi_n,\Phi_m\right\rangle _{\Lc^2}=\delta_{n,m}$, and that $N(x):=\sum_{n=0}^{N-1}|\Phi_n(x)|^2$ be strictly positive and finite, almost everywhere (a.e.) in $X$: $0<N(x)<\infty$, a.e. in $X$. Of course, boundedness of $N(x)$ is guaranteed whenever $N<\infty$, otherwise it is not. Let now $\Hil$ be an $N$-dimensional Hilbert space, with scalar product $\left\langle .,.\right\rangle _\Hil$, and with o.n. basis $\C=\{c_n, \,n=0,1,2,\ldots,N-1\}$: $\left\langle c_n,c_m\right\rangle _\Hil=\delta_{n,m}$.  In \cite{gazetal} it is proved that the vector $f_x:=\frac{1}{\sqrt{N(x)}}\,\sum_{n=0}^{N-1}\overline{\Phi_n(x)}\,c_n$, $x\in X$, has the following properties: (i) $\left\langle f_x,f_x\right\rangle _\Hil=1$, $\forall\, x\in\, X$; (ii) it satisfies the following resolution of the identity:
$$
\int_X\,|f_x\rangle\langle f_x|\,N(x)\,d\mu(x)=\1_N,
$$
where $\1_N$ is the identity operator in $\Hil_N$. This suggests that the set $\{f_x, \,x\in X\}$ can be efficiently used to quantize a given classical system, by going from a classical function $h(x)$ to its {\em upper symbol} $A_h:=\int_X\,h(x)\,|f_x\rangle\langle f_x|\,N(x)\,d\mu(x)$.

\vspace{2mm}

In view of what we have seen in the previous sections, it may be interesting to see if a similar strategy can be extended to the case when, rather than an o.n. basis, we have two biorthogonal bases. Notice also that we are mainly interested in considering finite-dimensional Hilbert spaces. As a matter of fact, this extension is quite straightforward: let $\Lc^2(X,d\mu)$ be as above, and let $\F_N^{(e)}:=\{e_n, \,n=0,1,2,\ldots,N-1\}$ and $\F_N^{(h)}=\{h_n, \,n=0,1,2,\ldots,N-1\}$ be two biorthogonal sets of the $N$-dimensional Hilbert space $\Hil_N$. Both sets are automatically bases for $\Hil_N$ and, in so far as $N<\infty$, they are biorthogonal Riesz bases: $\left\langle e_n,h_m\right\rangle _\Hil=\delta_{n,m}$ and they are the images, by means of a bounded operator $T$ with bounded inverse, of a certain o.n. basis of $\Hil_N$, $\C=\{c_n, \,n=0,1,2,\ldots,N-1\}$, $\left\langle c_n,c_m\right\rangle _\Hil=\delta_{n,m}$: $e_n=T\,c_n$ and $h_n=T^{-1}\,c_n$, $\forall n$. For symmetry reasons, it is natural to consider two biorthogonal sets $\{\Phi_n(x), \,x\in X: \,n=0,1,\ldots,N-1\}$ and $\{\Psi_n(x), \,x\in X: \,n=0,1,\ldots,N-1\}$ in $\Lc^2(X,d\mu)$, satisfying the following:
$$
0<\tilde N(x):=\sum_{n=0}^{N-1}\overline{\Phi_n(x) }\Psi_n(x)<\infty,
$$
a.e. in $X$. This apparently strange formula reduces to  the previous one if $\Phi_n(x)=\Psi_n(x)$ for all $n=0,1,2,\ldots,N-1$, or when these functions differ by some exponential term, from an o.n. set $\{\varphi_n(x), \,x\in X: \,n=0,1,\ldots,N-1\}$: $\Phi_n(x)=e^{\alpha_n(x)}\,\varphi_n(x)$ and $\Psi_n(x)=e^{-\alpha_n(x)}\,\varphi_n(x)$ for all $n=0,1,2,\ldots,N-1$, for rather general real functions $\alpha_n(x)$\footnote{Of course the main requirement is that these $\alpha_n(x)$ must be such that $\Phi_n(x)$, $\Psi_n(x)$ and $\varphi_n(x)$ all belong to $\Lc^2(X,d\mu)$.}.

The next step is to introduce two families of {\em coherent}-like states,
$$
e(x)=\frac{1}{\sqrt{\tilde N(x)}}\,\sum_{n=0}^{N-1}\,\Phi_n(x)\,e_n,\qquad h(x)=\frac{1}{\sqrt{\tilde N(x)}}\,\sum_{n=0}^{N-1}\,\Psi_n(x)\,h_n,
$$
extending what we have seen before. It is evident that $\left\langle e(x),h(x)\right\rangle _\Hil=1$ for a.a. $x\in X$, and that
$$
\int_X\,|e(x)\rangle\langle h(x)|\,\tilde N(x)\,d\mu(x)=\1_N.
$$
For this reason, these states are called bi-coherent states and, of course, they could be used to quantize each classical function, producing its upper symbol. Of course, if we were interested in producing eigenstates of some lowering operator, the situation changes drastically, since grassmann or paragrassmann variables are needed, \cite{bagpf} and \cite{gazetal}, with many extra difficulties. However, this is not our main interest here, and for this reason we are not considering this problem in this paper.

\vspace{2mm}

This procedure can be applied in each finite-dimensional Hilbert space of the type considered in Sections III. It turns out that in each $\Hil_M$ we can naturally introduce two sets of bi-coherent states, related by the operators $S_M^{(h)}$ and $S_M^{(e)}$. We could also extend, at least formally, most of the construction to $\Hil$ itself. In this case, we very luckily lose the boundedness of the intertwining operators, and a special (mathematical) care would be needed. Again, this is not our main interested here, and the analysis of this problem is postponed to a future paper.

\section{Conclusions}

We have shown how a certain noncommuting quantum mechanical system produces, in quite a natural way, a sequence of finite dimensional subspaces
of $\Lc^2({\Bbb R}^2)$ in which biorthogonal Riesz bases can be constructed explicitly,  which are eigenstates of two families of non
self-adjoint operators and of their adjoints.  They produce, under suitable conditions, two biorthogonal  bases of $\Lc^2({\Bbb R}^2)$ which
are not, in general, Riesz bases. This is done by working in a sequence of mutually orthogonal, finite dimensional, Hilbert spaces, into which
$\Lc^2({\Bbb R}^2)$ can be decomposed. Intertwining operators arising from this structure are also explicitly constructed. Examples in finite
dimensional vector spaces are discussed.

We have also shown how bicoherent states can
 be introduced in this settings, and we have discussed some of their properties.

\section*{Acknowledgements}
   The authors would like to acknowledge financial support from the
   Universit\`a di Palermo through Bando CORI, cap. B.U. 9.3.0002.0001.0001.
One of us (STA) would like to acknowledge a grant from the Natural Sciences and Engineering Research Council (NSERC) of Canada. The authors
also thank the referee for his very useful remarks.

\end{document}